# Improving the Renormalization Group approach to the quantum-mechanical double well potential

D. Zappalà[1]

*INFN, Sezione di Catania*
*Dipartimento di Fisica, Università di Catania*
*Corso Italia 57, I-95129, Catania, Italy*

## ABSTRACT

The gap between ground and first excited state of the quantum-mechanical double well is calculated using the Renormalization Group equations to the second order in the derivative expansion, obtained within a class of proper time regulators. Agreement with the exact results is obtained both in the strong and weak coupling regime.

Pacs 11.10.Hi ; 03.65.Ca

---

[1]E-mail address: dario.zappala@ct.infn.it

One of the most representative problems in quantum mechanics is the tunnelling of a particle through a potential barrier. It does not have a classical counterpart and cannot be handled in the usual perturbative approach. The double well potential provides a typical example of quantum tunnelling and, when employed in field theory, it also represents the most elementary toy model for the spontaneous symmetry breaking phenomenon and the related phase transitions and therefore it is extremely relevant in particle physics and in cosmology. The spectrum of the quantum mechanical double well can of course be evaluated numerically and it is known that the energy gap $\Delta E$ between the first excited state and the ground state, which appears as a consequence of the tunnelling between the two classical vacua, has a singular behavior in the quartic coupling $\lambda$: $\Delta E \propto \exp(-1/\lambda)$ [1]. The exponential behavior is well reproduced by the dilute istanton gas calculation, $\Delta E = 2\sqrt{2\sqrt{2}/(\pi\lambda)}\exp(-1/(3\sqrt{2}\lambda))$, which however is reliable only for very small values of $\lambda$ and becomes soon very different from the exact value of $\Delta E$ as $\lambda$ grows.

Some time ago it has been suggested [2, 3] that $\Delta E$ could be evaluated by making use of the Renormalization Group (RG). The original formulation of the non-perturbative RG obtained through a recursion procedure is due to Wilson [4] and since then a differential formulation of the RG flow equations has been developed [5]-[10]. The RG equations involve the full action of the problem considered and determine its flow as a function of a momentum scale $k$, starting from the bare action defined at a high momentum scale $k = \Lambda$ down to the infrared region $k \sim 0$. In the $k \to 0$ limit the running action contains the effects of all the fluctuations below the scale $\Lambda$ and can be identified with the effective action.

In [2, 3] the equation for the full action was approximated by an equation for the potential, which corresponds to the lowest order approximation in a systematic derivative expansion of the action. In [2] a further approximation was considered by expanding the potential in a polynomial series and then truncating this series. The flow equation for the potential was then reduced to a set of coupled ordinary differential equations for the various coefficients of the series. In [3] the full partial differential equation for the potential was instead studied with no further approximation. In both cases the results are practically the same: when $k$ is lowered the potential evolves from the bare double well shape, reducing the barrier until it totally disappears. When $k$ is close to zero the potential has become convex with the central region between the original minima almost flat. This is already an interesting feature because the convexity of the effective potential is an exact property [11] which can be observed by making use of non-perturbative approaches (see e.g. [12]) and cannot be recovered in perturbation theory. On a more quantitative level, however the flow equation for the potential is not fully satisfactory. In fact when it is used to compute the quantity $\Delta E$, which as explained in [2] is related to the second derivative of the effective potential at the origin, it provides accurate results for large values of the coupling $\lambda$, but when approaching $\lambda = 0$ the correct exponential behavior is not recovered. Therefore the regime in which the quantum tunnelling effects are relevant ($\lambda \sim 0$) is not well described neither by the approximation considered in [2] nor by the one in [3].

In this Letter we examine the natural improvement on the lowest order approximation of the flow equations in the derivative expansion, and specifically we include the next term in this expansion, i.e. the coefficient of the kinetic term of the action, $Z(k, x)$, which depends both on the running scale $k$ and on the spatial coordinate $x$. In field theory this is nothing else than a wave function renormalization and in quantum mechanics it has the role of a position dependent effective mass. We expect that the inclusion of this term, which represents the first correction to the fully local information carried by the potential could be helpful in describing the complex dynamics of the tunnelling.

Our aim is not only the analysis of a possible improvement in determining $\Delta E$, due to the inclusion of $Z(k, x)$. In fact we are particularly interested in checking the reliability of a particular version of the RG flow obtained by means of a regulator introduced in the Schwinger Proper Time (PT) formalism [13, 14, 15, 16]. As discussed in [16] and more in detail in [17] this particular flow, unlike the so called Exact Renormalization Group (ERG) flow, does not have a first principle derivation and there is no proof of convergence to the effective action in the $k \to 0$ limit. However, as shown in [15, 16], the PT RG equations are particularly interesting because they provide excellent determinations of the critical exponents at the non-gaussian fixed point in three dimensions, certainly comparable to the ones obtained with the ERG equations. The problem of determining $\Delta E$ in the quantum mechanical double well is therefore another good check for the PTRG because at least in this case the exact results are available.

In [15, 16] the flow equations of the PTRG are derived in the required approximation of the derivative expansion of the action, i.e. the coupled partial differential flow equations for the potential and for the wave function renormalization are displayed for the particular PT regulator $f_k(sZk^2) = \exp(-sZk^2) \sum_{i=0}^{m} (sZk^2)^i/i!$ where $s$ is the proper time and $m$ is an integer index. Actually the best values for the critical exponents in [15, 16] are obtained in the limit $m \to \infty$ and therefore here we are particularly interested in this limit. In [16] the limit $m \to \infty$ is formally taken in the fixed point equations by redefining the field variables. It is also noticed that the second derivative of the potential determined by the $m$-dependent equations behaves like $1/m$ for large $m$. This behavior would imply in the present case that $\Delta E = 0$ in the large $m$ limit. However this problem can be avoided if one, instead of considering the field redefinition as in [16], would simply start with a slightly different cut-off function $\widetilde{f}_k(sZk^2) = \exp(-sZmk^2)\sum_{i=0}^{m}(sZmk^2)^i/i!$ where $k^2$ has been replaced by $mk^2$. This is practically equivalent to what has been done in [17] where the same rescaling of $k$ has been performed directly in the flow equations. However the latter approach mixes the two limits $m \to \infty$ and $k \to 0$ and this could generate some confusion.

If the new cut-off function $\widetilde{f}_k$ is used instead of $f_k$, a new set of $m$-dependent flow equations are obtained whose large $m$ limit (this time with no redefinition of the field or of the scale $k$) yields again the equations which have been determined in [16] and [17]. The $m$-dependent flow equations obtained with the cut-off $\widetilde{f}_k$ as we shall see below provide for the double well problem, in the limit $m \to \infty$ a non-vanishing second derivative of the potential and therefore a finite value of $\Delta E$. The procedure outlined in [15] to derive the flow equations, with the cut-off $f_k$ replaced by $\widetilde{f}_k$, yields the two coupled partial differential equations for the running potential $V(k,x)$ and the wave function renormalization $Z(k,x)$

$$k\frac{\partial V}{\partial k} = \alpha(k^2 m)^{D/2}\left(\frac{Zk^2}{Zk^2 + V''/m}\right)^{m+1-D/2} \tag{1}$$

$$k\frac{\partial Z}{\partial k} = \alpha(k^2 m)^{D/2}\left(\frac{Zk^2}{Zk^2 + V''/m}\right)^{m+1-D/2}\left[\frac{(m+1-D/2)}{m(Zk^2+V''/m)}\left(-Z''\right.\right.$$
$$\left.+\frac{(4+18D-D^2)(Z')^2}{24Z}\right) + \frac{(10-D)(m+1-D/2)(m+2-D/2)}{6m^2(Zk^2+V''/m)^2}Z'V'''$$
$$\left.-\frac{(m+1-D/2)(m+2-D/2)(m+3-D/2)}{6m^3(Zk^2+V''/m)^3}Z(V''')^2\right] \tag{2}$$

where each prime indicates a derivative w.r.t. the spatial coordinate $x$ and the constant $\alpha$ is expressed in terms of gamma functions

$$\alpha = \frac{\Gamma(m+1-D/2)}{(4\pi)^{D/2}\Gamma(m+1)} \tag{3}$$

Eqs. (1,2) in the limit $m \to \infty$ become

$$k\frac{\partial V}{\partial k} = \left(\frac{k^2}{4\pi}\right)^{D/2} e^{-V''/(Zk^2)} \tag{4}$$

$$k\frac{\partial Z}{\partial k} = \left(\frac{k^2}{4\pi}\right)^{D/2} e^{-V''/(Zk^2)}$$
$$\times\left(-\frac{Z''}{Zk^2} + \frac{(4+18D-D^2)(Z')^2}{24Z^2 k^2} + \frac{(10-D)Z'V'''}{6(Zk^2)^2} - \frac{Z(V''')^2}{6(Zk^2)^3}\right) \tag{5}$$

Note that all equations are expressed in terms of dimensionful quantities (with $\hbar = 1$). The determination of $\Delta E$ requires the numerical integration of the coupled equations for $V$ and $Z$ with $D = 1$. We have to fix the initial conditions of the differential equations which we take at a value $k = \Lambda$, with $\Lambda$ much larger than any other scale in the problem. The double well potential

$$V_{dw}(x) = \frac{1}{2} M^2 x^2 + \lambda x^4 \tag{6}$$

corresponds to the initial condition of the flow of the potential $V(k, x)$ whereas, according to the normalization of the kinetic term in the bare action, we take $Z = 1$ as initial condition for $Z(k, x)$. The scale of all the dimensionful quantities is fixed by the choice $M^2 = -1$. We shall also consider, for comparison, the convex anharmonic oscillator, and in this case we choose $M^2 = 1$. Correspondingly the scale at which the initial conditions of the parameters are fixed is chosen $\Lambda = 1500$. The numerical resolution of the two coupled partial differential equations is performed with the help of the NAG routines.

In Fig. 1 $V''(0,0)$ and $Z(0,0)$ obtained from Eqs. (1,2) are reported versus $1/m$. Namely black circles and diamonds correspond respectively to $Z(0,0)$ and $V(0,0)$ for $M^2 = 1$ and $\lambda = 0.4$. White circles and diamonds correspond to $Z(0,0)$ and $V(0,0)$ for $M^2 = -1$ and $\lambda = 0.05$. In all cases we have inserted at $1/m = 0$ the corresponding values obtained from Eqs. (4,5). The convergence to these latter values is clear. Therefore, as we had anticipated, in the limit $m \to \infty$, Eqs.(1,2) provide non-vanishing values of the second derivative of the potential which converge to those obtained from Eqs.(4,5).

Since we are particularly interested in the $1/m = 0$ case, from now on we consider Eqs.(4,5). The qualitative behavior of the running potential along the flow is explained in detail in [3] and we observe a similar trend. In Fig. 2 the second derivative of the running potential (continuous lines) and the wave function renormalization (dashed lines) obtained at various values of $k$ (namely $k = 1500, 10, 1, 0.5, 0.2, 0.1, 0$) for $M^2 = -1$ and $\lambda = 0.06$ are plotted. Below $k = 0.2$ the second derivative of the potential at the origin becomes positive and the potential convex. At the same time $Z$, which for most of the running stays close to one, rapidly increases in the region between the classical minima with a sharp fall to one outside that region. The highest peak corresponds to the final output at $k = 0$.

Finally we turn to the evaluation of the energy gap $\Delta E$. As explained in [2], as long as the potential is analysed keeping $Z = 1$ fixed during the flow, $\Delta E$ is to be identified with $\sqrt{V''(0,0)}$, but in the approximation here considered with a running wave function renormalization, the energy gap $\Delta E$ corresponds, in field theory language, to the renormalized mass, i.e.

$$\Delta E = \sqrt{V''(0,0)/Z(0,0)} \qquad (7)$$

Therefore we determined the values of $\Delta E$ from Eqs.(4,5) according to Eq. (7) (these results are indicated as $\Delta E_{PTnlo}$) and, for comparison, we also evaluated $\Delta E$ to the lowest order in the derivative expansion (indicated as $\Delta E_{PTlo}$), i.e. solving the flow in Eqs.(4,5) keeping the wave function renormalization constant $Z = 1$. The energy gap to the lowest order in the derivative expansion with the flow equation used in [3], which is identical to the Wegner-Houghton local potential equation in $D = 1$ (see [5])

$$k\frac{\partial V}{\partial k} = \frac{-k}{2\pi}\log\left(1 + \frac{V''}{k^2}\right) \qquad (8)$$

has also been calculated ($\Delta E_{wh}$) and all these results are eventually compared with the exact estimate $\Delta E_{exact}$, derived by numerically solving the quantum mechanical eigenvalue problem. In Table 1 these estimates of $\Delta E$ are collected for various values of the coupling $\lambda$ both for $M^2 = 1$ and $M^2 = -1$. The values found for $Z(0,0)$ are also included. From the accuracy of the numerical code the estimated errors in Table 1 are at most two or three units on the last digit.

In the $M^2 = 1$ case, as expected from [2, 3], everything is rather smooth. In fact we observe that the differences among the various determinations of $\Delta E$ are extremely small and $Z(0,0)$ is practically equal to its bare value $Z(\Lambda, x) = 1$. However little discrepancies appear at large values of $\lambda$ (strong coupling regime) and $\Delta E_{wh}$ turns out to be a slightly better approximation than $\Delta E_{PTlo}$ and to be comparable to $\Delta E_{PTnlo}$.

The double well problem with $M^2 = -1$ is rather different. Here, from [2, 3] we know that the RG equations work better in the strong coupling regime. Actually this feature is true for $\Delta E_{wh}$ which, in the region where the tunnelling plays an important role ($\lambda < 0.1$), does not reproduce the exponential behavior of $\Delta E_{exact}$. On the other side, $\Delta E_{PTlo}$ which is less accurate at $\lambda = 0.4$, turns out to be more reliable at small values of $\lambda$. But the remarkable result is that the higher order approximation is very accurate. In fact the correction due to the wave function renormalization increases when $\lambda$ is decreased and the agreement between $\Delta E_{PTnlo}$ and $\Delta E_{exact}$ is always excellent. Unfortunately below $\lambda = 0.05$ the peak in $Z(0,x)$ shown in Fig. 2 becomes very narrow, generating problems in the numerical resolution of the flow equations and we were not able to determine the corresponding values of $Z(0,0)$ and of the energy gap. In any case for such small values of $\lambda$ the numerical errors are no longer negligible and a more refined analysis should be performed. Even an approximated analytical investigation of this regime, similar to the one discussed in [3], would be useful, although the problem of two coupled equations is much more difficult to handle.

In conclusion we have determined the energy gap between the ground state and the first excited state of the quantum mechanical potential $V_{dw}$ in Eq. (6) both for positive (quartic anharmonic potential) and for negative (double well potential) curvature $M^2$, by making use of the RG techniques. In particular we used a Proper Time regulated version of the RG flow equations in the derivative expansion approximation, truncated to the second order. The results show the reliability of the PT RG flow which, even to the lowest order provides good estimates of the energy gap and, above all, they show the sensible improvement coming from the inclusion of the wave function renormalization in the quantitative analysis of a tunnelling process.

**Acknowledgement**

The author is grateful to Alfio Bonanno for many fruitful discussions.

| $\lambda$ | $\Delta E_{wh}$ | $\Delta E_{PTlo}$ | $\Delta E_{exact}$ | $\Delta E_{PTnlo}$ | $Z(0,0)$ |
|---|---|---|---|---|---|
| $M^2 = 1$ | | | | | |
| 1.0 | 1.9291 | 1.9464 | 1.9341 | 1.9380 | 1.0052 |
| 0.4 | 1.5450 | 1.5556 | 1.5482 | 1.5498 | 1.0037 |
| 0.1 | 1.2091 | 1.2127 | 1.2104 | 1.2109 | 1.0013 |
| 0.05 | 1.1201 | 1.1218 | 1.1208 | 1.1210 | 1.0006 |
| 0.03 | 1.0774 | 1.0784 | 1.0779 | 1.0780 | 1.0003 |
| 0.02 | 1.0538 | 1.0544 | 1.0540 | 1.0542 | 1.0002 |
| $M^2 = -1$ | | | | | |
| 0.4 | 0.9654 | 0.9897 | 0.9667 | 0.9730 | 1.0217 |
| 0.3 | 0.8173 | 0.8404 | 0.8166 | 0.8233 | 1.0273 |
| 0.2 | 0.6212 | 0.6416 | 0.6159 | 0.6227 | 1.0416 |
| 0.1 | 0.3297 | 0.3280 | 0.2969 | 0.3027 | 1.1321 |
| 0.07 | 0.2238 | 0.1848 | 0.1539 | 0.1562 | 1.3343 |
| 0.06 | 0.1902 | 0.1311 | 0.1031 | 0.1028 | 1.5548 |
| 0.05 | 0.1576 | 0.0806 | 0.0562 | 0.0532 | 2.1270 |
| 0.04 | 0.1259 | 0.0496 | 0.0210 | $----$ | $----$ |
| 0.03 | 0.0947 | 0.0329 | 0.0036 | $----$ | $----$ |
| 0.02 | 0.0637 | 0.0204 | 0.0003 | $----$ | $----$ |

Table 1: Various determinations of $\Delta E$, including its exact estimate, and of $Z(0,0)$ for some values of the coupling $\lambda$, both in the $M^2 = 1$ and in the $M^2 = -1$ case.

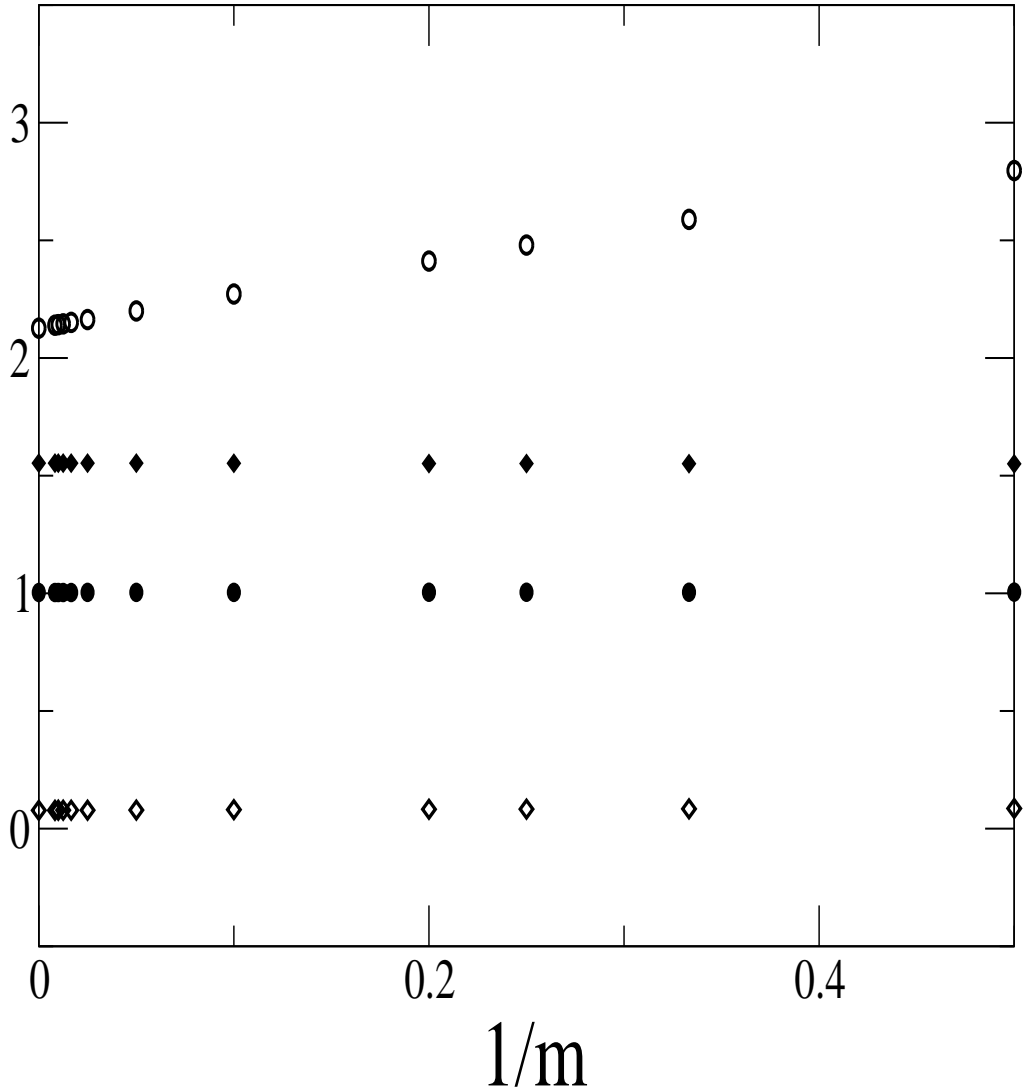

Figure 1: $V''(0,0)$ ( black and white diamonds) and $Z(0,0)$ (black and white circles) obtained from Eqs.(1,2) respectively for $M^2 = 1$, $\lambda = 0.4$ and for $M^2 = -1$, $\lambda = 0.05$, vs $1/m$. The limiting values at $1/m = 0$ are obtained from Eqs. (4,5).

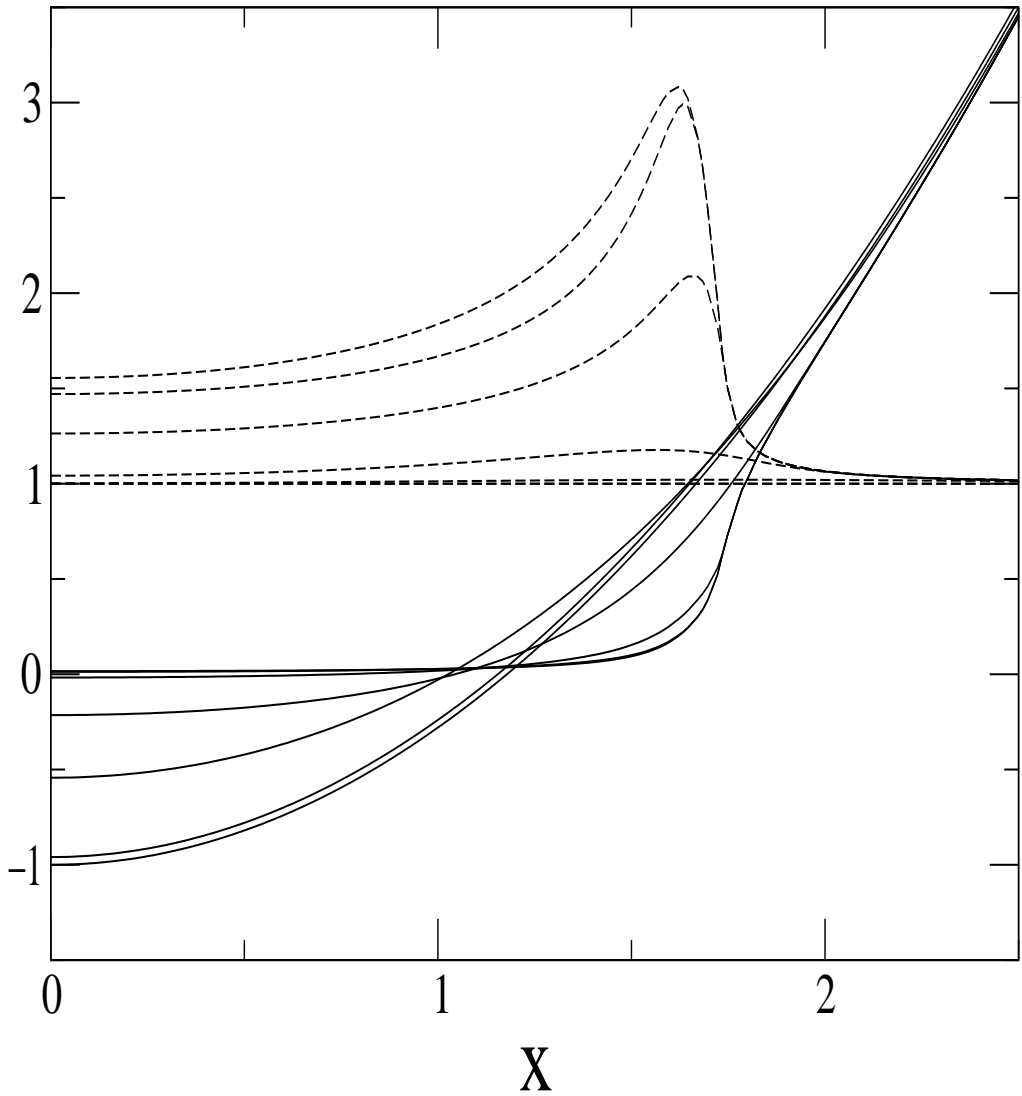

Figure 2: $V''(k, x)$ (continuous lines) and $Z(k, x)$ (dashed lines) plotted *vs* $x$ at $k = 1500, 10, 1, 0.5, 0.2, 0.1, 0$, as obtained from Eqs. (4,5) for $M^2 = -1$ and $\lambda = 0.06$.

10